\documentclass[fleqn,twoside]{article}
\usepackage[headings]{espcrc2}

\readRCS
$Id: espcrc2.tex,v 1.2 2004/02/24 11:22:11 spepping Exp $
\usepackage{graphicx}
\usepackage[figuresright]{rotating}

\newcommand{\noun}[1]{\textsc{#1}}

\newcommand{\AmS}{{\protect\the\textfont2
  A\kern-.1667em\lower.5ex\hbox{M}\kern-.125emS}}

\title{Hadronic Interactions at Cosmic Ray Energies}

\author{S. Ostapchenko\address{Forschungszentrum Karlsruhe, Institut f\"ur Kernphysik, 76021 Karlsruhe, Germany}%
    \address{D.V. Skobeltsyn Institute of Nuclear Physics, Moscow
State University, 119992 Moscow, Russia}}
       
\runtitle{Hadronic Interactions at Cosmic Ray Energies}
\runauthor{S. Ostapchenko}

\begin{document}

\begin{abstract}
General physics of very high energy	hadronic interactions is discussed.
Special attention is payed to the contribution of semihard
processes to the interaction dynamics and to the role of parton shadowing and
parton density saturation. In particular, the implementation of non-linear
interaction effects in the \noun{QGSJET-II} model is discussed in detail.  The predictions of the model are compared to selected accelerator data,
 including ones of the RHIC collider, and the relation to the calculated 
extensive air shower characteristics is discussed. 
Finally, the potential of  accelerator and cosmic ray  experiments for constraining  model predictions  is analyzed.
\vspace{1pc}
\end{abstract}

\maketitle

\section{INTRODUCTION}
During the last decade a significant progress has been achieved in experimental studies of high
energy cosmic rays (CR) with extensive air shower (EAS) techniques, both concerning the 
measurements of the primary cosmic ray (PCR) energy spectrum and in determining the
 composition of cosmic radiation in the region of the spectral
``knee'' and at higher energies. This resulted from further development of the 
measurement and data analysis techniques,  the latter including thorough EAS
simulation studies with contemporary  Monte Carlo (MC) tools.
 In EAS calculations a special role is played by hadronic MC generators
which are used for the description of hadron-air and nucleus-air interactions in air showers,
performing an extrapolation of current theoretical and experimental knowledge 
 towards the highest CR energies. In particular, the \noun{QGSJET}
model \cite{kal94,kal97}, being based on the Gribov's effective Reggeon Field Theory (RFT) \cite{gri68,gri69,bak76,kai79} and the Pomeron phenomenology
 \cite{kai84,cap94}, proved to be very  
 successful in describing air shower data obtained by various experimental installations.
The original  versions of this MC generator  have been 
developed as MC realizations of the Quark-Gluon String (QGS)
 model \cite{kai84}, including
a generalization of the QGS model approach for the treatment of nucleus-nucleus collisions
and a description of the fragmentation of nuclear spectator part \cite{kal89,kal93}.
In  \noun{QGSJET} this has been supplemented by a  phenomenological treatment of 
 semihard processes, which result in the production of observed hadron jets of comparatively 
 high transverse momenta. Finally, in the \noun{QGSJET-II} model \cite{ost06c}
 a treatment of non-linear interaction effects has been developed, 
based on all-order  re-summation of  so-called enhanced (Pomeron-Pomeron
 interaction) RFT diagrams \cite{ost06a,ost06b}. 
These approaches will be discussed below, in comparison with other models 
and with selected accelerator data.

\section{MODEL FRAMEWORK} \label{frame}
\subsection{Basic physics} \label{basic}
The general picture for high energy hadronic collision is the one of a multiple
scattering process, being mediated by multiple parton (quark and gluon)
cascades proceeding between the two hadrons. In the center of mass
frame such parton cascades develop on a much larger time scale than the one
of a parton-parton scattering, so that in the moment of the collision
 both projectile and target hadrons are represented by their parton clouds. 
Binary re-scatterings of some of these partons violate the coherence of the
 corresponding parton chains (consisting of the parton ``parents'' and ``pre-parents''), which then fragment into secondary hadrons. 
Alternatively,
the coherence may be preserved for some parton re-scatterings, in which case
all partons from the corresponding chains recombine back to their
parent hadrons without a production of secondaries, which gives rise to 
elastic re-scattering processes. A general inelastic collision involves a 
number of elementary hadron production contributions as well as multiple
elastic re-scatterings. In turn, elastic scattering is obtained when only
elastic sub-processes occur.

\begin{figure}[t]
\vspace{4pt}
\begin{center}
\includegraphics*[width=6.5cm,height=2.cm,angle=0]{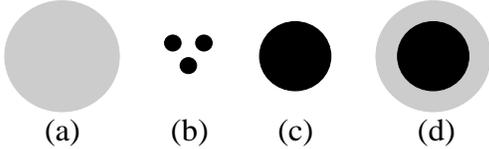}
\vspace{-.8cm}
\caption{Proton profile as viewed in soft (a), hard (b), semihard (c), and
 general (d) interactions at very high energies.}
\label{fig:hadron-profile}
\end{center}
\vspace{-0.8cm}
\end{figure}

With the energy increasing, the number of elementary re-scattering processes
 grows rapidly, due to the larger phase space for parton
emissions. In addition, one expects a qualitative change in the structure of the
underlying parton cascades. Indeed, at comparatively low energies all the
 partons are characterized by  small transverse momenta; high 
$p_t$ emissions are suppressed by the smallness of the corresponding running
coupling, $\alpha _s(p_t^2)$. By the uncertainty principle,   each 
emission  is characterized by  a large displacement of the produced parton in the transverse plane, $\Delta b^2\sim 1/p_t^2$.
 Thus, with the energy increasing further,
such ``soft'' parton cascades rapidly expand towards larger impact parameters,
 while the density of partons per unit transverse area remains small, the
hadron looking ``grey'', as depicted in Fig.~\ref{fig:hadron-profile}~(a).
 However, at sufficiently high energies an important contribution comes
from so-called  ``semi-hard'' and ``hard'' parton cascades, in which some
or all partons have comparatively high transverse momenta \cite{glr}. There,
the smallness of the strong coupling  $\alpha _s(p_t^2)$ is compensated
by a high parton density and by large logarithmic ratios of  the longitudinal
and transverse momenta for successive parton emissions. Purely hard cascades,
which start, e.g., from  valence quarks and contain only high $p_t$ partons,
do not expand  transversely,  $\Delta b^2\sim 1/p_t^2$ being small,
and lead to an increase of parton density in small areas (``hot spots'')
in the transverse plane -- see Fig.~\ref{fig:hadron-profile}~(b), while giving
a negligible contribution to the total cross section. Contrary to that,
typical semihard re-scatterings are two-step processes: first, parton
 branchings proceed with a small momentum transfer and the cascade develops
towards larger impact parameters; next, high $p_t$ parton emissions become
effective, leading to a rapid rise of the parton density at a given
point in the transverse plane. As a result, the region of high parton density
extends to large impact parameters (Fig.~\ref{fig:hadron-profile}~(c)) and
the contribution dominates in the very high energy limit. General hadronic
interactions include all the mentioned mechanisms; hadrons in high energy
collisions look as shown in Fig.~\ref{fig:hadron-profile}~(d): there is an
extended ``black'' region of high density, dominated by the semihard processes,
and around it there is a ``grey'' region of low  density, formed by
purely soft parton cascading \cite{str06}. In the ``black'' region one
expects strong non-linear parton effects to emerge, which result in the
saturation of parton densities and in the suppression of soft parton emissions
\cite{glr}.  On the other hand, such effects are negligible
 in the ``dilute'' peripheral region.

How to estimate the relative importance of the two regimes?  
Small  peripheral contribution would correspond to the ``black
disc'' limit for  hadron-hadron scattering, with the ratio of elastic to total
cross sections approaching 1/2. The relative smallness of the observed 
 $\sigma ^{\rm el}_{pp}/ \sigma ^{\rm tot}_{pp}$ ratio
indicates that the ``black''
 central region and the peripheral one are yet of comparable sizes \cite{str06}.

In reality, the discussed separation of central and peripheral collisions
 is rather crude, as the average parton 
densities rise gradually with decreasing impact parameter.
 Thus, there exists an important ``transition'' region of
moderately large impact parameters, characterized by large but not yet
 saturated parton densities, where the contributions of both soft and semihard processes 
 are of equal importance, and where non-linear parton
effects  provide sizable corrections.
In fact, it is this  transition region which is expected to give
the dominant contribution to  the interaction characteristics relevant for EAS physics.

\subsection{Pomeron formalism} \label{pomform}
\begin{figure}[t]
\vspace{4pt}
\begin{center}
\includegraphics*[width=6cm,height=2.25cm,angle=0]{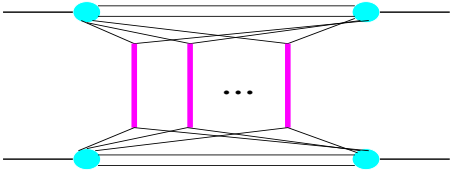}
\vspace{-8mm}
\caption{ A general contribution to hadron-hadron
 scattering amplitude. Elementary scattering processes (vertical thick lines)
are described as Pomeron exchanges. \label{multi-pom}}
\vspace{-0.8cm}
\end{center}
\end{figure}

As discussed above, general hadronic collisions necessarily involve
emissions of soft low $p_t$ partons, which prevents one from applying
the perturbative QCD formalism. However, for a number of key characteristics, 
like total and elastic cross sections or probabilities of various
``macroscopic'' configurations of inelastic interactions, the knowledge
of the microscopic parton picture is not of extreme importance; the RFT
 allows one to calculate the quantities
of interest based on the knowledge of the elastic amplitude for an 
``elementary'' scattering process. For example, in the QGS model \cite{kai84}
 hadron-hadron scattering is described as a multiple exchange of
composite objects -- Pomerons, as shown in Fig.~\ref{multi-pom}.
Each Pomeron represents  a microscopic parton cascade whose precise
 description is not necessary at this stage. Rather one employs the 
theoretically-motivated ansatz for the corresponding scattering amplitude:
\begin{equation}
 f^{\rm P}_{ad}\! \left(s,b\right) =
  \frac{i\gamma_{a}\gamma_{d}\,s^{\alpha_{\rm P}(0)-1}\,
e^{-\frac{b^{2}} {4 ( R_{a}^{2}+R_{d}^{2}  + \alpha_{\rm P}'(0)\ln s)}}}
{R_{a}^{2}+R_{d}^{2}  + \alpha_{\rm P}'(0)\ln s},
\end{equation}
which is characterized by a power-like energy rise and  by a logarithmically
increasing slope; the parameters $\alpha_{\rm P}(0)$, $\alpha_{\rm P}'(0)$ 
are the intercept and the slope of the Pomeron Regge trajectory and 
  $\gamma _a$, $R^2_a$ -- the residue and the slope for the  Pomeron-hadron $a$ 
  vertex \cite{kai84,cap94}.%

It is important to take into account the contributions of both elastic and 
inelastic intermediate hadron states between Pomeron emissions, such that 
the Pomeron-hadron  vertex becomes a matrix corresponding to the transitions
between those states. To diagonalize this matrix one considers any hadron to be 
represented by a superposition of a number of elastic scattering eigenstates,
$|a\rangle = \sum _j \sqrt{C_a^{(j)}}\, |j_a \rangle$, which are characterized by 
different absorption in the scattering process,
 $\gamma_{a}^{(j)}=\lambda_a^{(j)}\, \gamma_{a}$ \cite{kai79}.
 Here $C_a^{(j)}$ and $\lambda_a^{(j)}$ are the weights of the
 eigenstates and the relative strengths
for their coupling to the Pomeron; $\sum _j C_a^{(j)}=1$,
 $\sum _j C_a^{(j)}\,\lambda_a^{(j)}=1$.
Then, the total cross section is obtained as the sum of partial
scattering contribitions, being expressed via the so-called Pomeron
eikonal $\chi_{ad}^{\rm P}(s,b)={\rm Im}f^{\rm P}_{ad}\! \left(s,b\right)$
\begin{equation}
 \sigma_{ad}^{{\rm tot}}\!(s) = 2\!\! \int\!\!\! d^{2}b \sum_{j,k}C_a^{(j)}  C_d^{(k)} 
(1-e^{-\lambda_a^{(j)}\lambda_d^{(k)}
\chi_{ad}^{\rm P}(s,b)}) \label{sig-tot}
\end{equation}

\begin{figure}[t]
\vspace{4pt}
\begin{center}
\includegraphics*[width=6cm,height=2.25cm,angle=0]{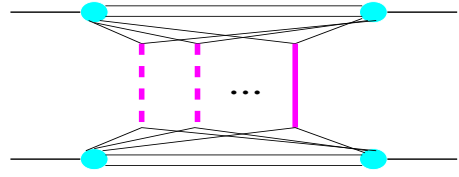}
\vspace{-8mm}
\caption{ Typical inelastic interaction contains a number of
 elementary  production processes, described by cut Pomerons 
 (thick broken lines in the Figure), 
and any number of elastic re-scatterings --
uncut Pomeron exchanges. \label{multi-cut}}
\vspace{-0.8cm}
\end{center}
\end{figure}

To obtain  cross sections for various final states,
one makes use of the optical theorem, which relates the 
total sum of contributions of all final states to the imaginary part
of the elastic  amplitude for hadron-hadron scattering, hence, to the
contributions of various unitarity cuts of  elastic scattering diagrams.
The so-called Abramovskii-Gribov-Kancheli  cutting rules \cite{agk}
 state that
only certain classes of cut diagrams are important in the high energy limit
and allow one to relate such contributions to particular final states of interest.
For example, the cross sections for having $n$ simultaneous elementary 
 production processes are described by the diagrams with $n$ ``cut''
Pomerons and any number of uncut ones, as shown in  Fig.~\ref{multi-cut},
whereas elastic and diffractive
cross sections are obtained cutting the diagrams of Fig.~\ref{multi-pom}
between the Pomerons, with no one being cut, and selecting elastic 
or diffractive intermediate hadron states in the cut plane. Thus,
the cross sections for the $n$ cut Pomerons process and for 
the projectile  diffraction dissociation read
\begin{eqnarray}
\sigma_{ad}^{(n)}\!(s)=\int\!\! d^{2}b \sum_{j,k}C_a^{(j)} C_d^{(k)}
 && \nonumber \\
\times \;\frac{\left[ 2\lambda_a^{(j)}\lambda_d^{(k)}
\chi_{ad}^{\rm P}\!(s,b)\right]^n}{n!} 
\;  e^{-2\lambda_a^{(j)}\lambda_d^{(k)}
\chi_{ad}^{\rm P}\!(s,b)} && \label{sigma-(n)} \\
\sigma_{ad}^{{\rm DD}_a}\!(s)= \int\!\! d^{2}b \! \sum_{j,k,l,m} \!
\left(C_a^{(j)}\delta_{j}^{l}-C_a^{(j)}C_a^{(l)}\right)&& \nonumber \\
\times \;C_d^{(k)} C_d^{(m)}\;
e^{-(\lambda_a^{(j)}\lambda_d^{(k)}+\lambda_a^{(l)}\lambda_d^{(m)})
\,\chi_{ad}^{\rm P}\!(s,b)} &&
\label{LMD-proj}
\end{eqnarray}
From (\ref{LMD-proj}) follows that   a higher
diffraction cross section is obtained for  more asymmetric couplings 
$\lambda_{a(j)}$  of different eigenstates to the Pomeron. In turn,
this results in a smaller total cross section and larger fluctuations
in the number $n$ of elementary production processes (cut Pomerons),
as one can see from (\ref{sig-tot}--\ref{sigma-(n)}).

Having obtained partial probabilities of various inelastic final states,
one has yet to describe particle production for each elementary inelastic 
process, which leads us back to the physics of the underlying parton cascade.
However,  one can employ here the string picture of the hadronization,
based on the color structure of the corresponding final states. Namely,
one assumes that each cut Pomeron process induces a color exchange between
parton constituents of the interacting hadrons, such that color strings
are stretched between them. With the two hadrons flying apart, these strings
break up and hadronize, which is described by string fragmentation procedures. 
It is noteworthy, that in the QGS model the  parameters for constituent
 parton momentum distributions and the ones for
string fragmentation are not adjustable ones, being expressed via the
 characteristics of known Regge trajectories \cite{kai84,kai87}.

Remarkably, the described scheme can be generalized to hadron-nucleus and
nucleus-nucleus interactions in a parameter-free way, both for cross section
calculations and concerning the description of particle production 
\cite{kal93}. The only new input are nuclear density profiles,
 with the corresponding parameters being fixed by  nuclear form factor 
 measurements \cite{nucdat}, individually for each nuclear type, as discussed in \cite{kal99}.

\subsection{Semihard processes}
\begin{figure}[t]
\vspace{4pt}
\begin{center}
\includegraphics*[width=7.5cm,height=2.8cm,angle=0]{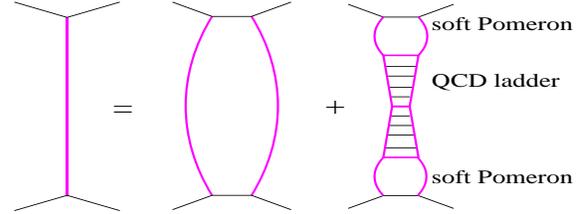}
\vspace{-8mm}
\caption{A general Pomeron is the sum of the ``soft'' and 
the ``semihard'' ones -- correspondingly the 1st and the 2nd graphs
in the r.h.s. \label{pomgen}}
\vspace{-0.8cm}
\end{center}
\end{figure}

As discussed in Section \ref{basic}, at high enough energies a significant
contribution to hadron-hadron scattering comes from semihard parton processes,
where a part of the underlying parton cascade develops in the high $p_t$
region. To provide a microscopic treatment of the corresponding physics,
one can apply the phenomenological Pomeron description to the low virtuality
($|q^2|\simeq p_t^2<Q^2_0$) part of the parton cascade, while treating the
high $p_t$  parton evolution using pQCD techniques, $Q^2_0$ being some
chosen virtuality cutoff for QCD being applicable. Then, an elementary 
scattering process is described as an exchange of a ``general Pomeron'',
which is  a sum of the ``soft'' and the ``semihard'' ones,  
the latter being represented by a piece of   QCD ladder sandwiched
between two ``soft'' Pomerons \cite{kal94,kal97,dre99,ost02}, as shown in Fig.~\ref{pomgen}.%
 As discussed in \cite{dre99,ost02}, the upper and the lower half of the semihard
 Pomeron define parton (sea quark and gluon) momentum and impact parameter
distributions in the projectile and target hadrons; the
 parameters for the coupling between the soft Pomeron and the ladder
can be fixed by the data on hadron structure functions (SFs).
The approach allows one to  treat  the semihard processes within the Pomeron
scheme described in Section \ref{pomform}; the interaction cross sections
are defined by the usual formulas, like (\ref{sig-tot}--\ref{LMD-proj}),
with the Pomeron eikonal being a sum of the soft and the semihard ones,
$\chi_{ad}^{\rm P}(s,b)=\chi_{ad}^{{\rm P}_{\rm soft}}(s,b)+
\chi_{ad}^{{\rm P}_{\rm sh}}(s,b)$. In turn,  particle production
procedure includes an explicit treatment of the high $p_t$ parton cascade,
with the strings being formed between the produced final partons
\cite{kal97,dre99,ost02}.

\subsection{Non-linear effects}\label{nonlinear}
Developing a model for high energy hadronic interactions, one inevitably faces
the problem of treating non-linear  effects, connected to parton shadowing
and saturation. Indeed, describing the interaction as a superposition of a 
number of re-scattering processes, mediated by parton cascades,
one has to consider the case when such
cascades overlap in the corresponding phase space and influence each other.
Such effects are expected to be extremely important at very high energies
and small impact parameters, i.e.~in the ``black'' region of high parton
 densities, where they lead to the parton density saturation \cite{glr}
and to significant reduction of secondary particle production. However, 
non-linear  effects start to be efficient already at comparatively
low energies and large impact parameters, the experimental indication
being the rapid energy rise of the high mass diffraction cross section 
in the ISR energy range \cite{goul}, the latter being just one of a number
 of manifestations of non-linear parton dynamics. Therefore, 
one has to develop a coherent description of the corresponding physics 
over a wide dynamic range,  rather than restrict himself with a 
treatment of the saturation region.

Treating independent parton cascades effectively as Pomeron exchanges, the
corresponding non-linear effects are described  in the RFT as Pomeron-Pomeron
interactions \cite{kan73,car74,cap76}. There, the main technical difficulty
is to perform a re-summation of contributions of
the underlying  enhanced Pomeron graphs, 
as  more and more diagrams of compicated
topologies come into play at higher energies. 
A re-summation method has been worked out
recently \cite{ost06a,ost06b} and implemented in the \noun{QGSJET-II} model
\cite{ost06c}. The basic assumption of the approach was that Pomeron-Pomeron
coupling proceeds  via parton processes at comparatively small
virtualities $|q^2|<Q^2_0$ and can be described using phenomenological
multi-Pomeron vertices of eikonal type \cite{ost06b}. A reasonable consistency
 with relevant experimental data has been obtained
using a fixed energy-independent $Q_0$-cutoff,\footnote{The parameter set
in \cite{ost06b} is different from the default  \noun{QGSJET-II} settings.
While a general agreement with measured proton-proton cross section and
proton SFs can be obtained for $Q^2_0$ as small as 1 GeV$^2$, the description
of particle production requires a higher cutoff, the default \noun{QGSJET-II}
value being 2.5 GeV$^2$. This may be considered as an indication for 
significant parton shadowing in the virtuality range $1\div 2.5$ GeV$^2$.}
neglecting parton shadowing effects for $|q^2|>Q^2_0$.
 In particular, inferring the basic parameter of the scheme, the triple-Pomeron
coupling, from HERA hard diffraction data, one obtains a satisfactory agreement
with observed hadronic diffraction \cite{ost06b}.

To describe particle production, one has to consider unitarity cuts of enhanced
Pomeron diagrams and, for any configuration of cut Pomerons, to perform a full
re-summation of uncut ones \cite{ost06b,ost07}. The big advantage of the 
developed procedure is that the solutions are obtained in the 
form of recursive equations, which can be implemented in a MC model and 
 allow one to generate various  configurations of interactions,
including diffractive ones, in an iterative fashion \cite{ost07}.

\begin{figure}[t]
\vspace{4pt}
\begin{center}
\includegraphics*[width=7.cm,height=4.7cm,angle=0]{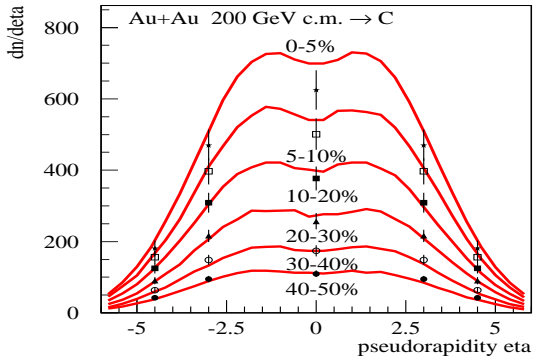}
\vspace{-8mm}
\caption{Pseudorapidity distributions of charged secondary particles
in Au-Au collisions of different ``centralities''; \noun{QGSJET-II}
results (lines) are compared to  BRAHMS data \cite{brahms}. \label{rhicn}}
\vspace{-0.8cm}
\end{center}
\end{figure}

The  formalism seems to be adequate for the description of hadronic
collisions in the  peripheral and ``transition'' regimes.
 However, in the ``black'' region of  high parton densities
 one may expect an important contribution of
 the  ``hard'' ($|q^2|>Q^2_0$) Pomeron-Pomeron coupling, which 
would provide additional screening corrections. A reasonable agreement of 
 \noun{QGSJET-II}  with RHIC data on central heavy ion collisions 
 (see Fig.~\ref{rhicn})%
 indicates the smallness of such effects. However, the
situation may change at much higher energies.

\subsection{Nuclear fragmentation}
Performing simulations of nucleus-induced air showers, one has to pay
attention to the fragmentation of nuclear spectator part, as it has a
significant influence on the predicted EAS fluctuations  \cite{kal93}. 
Nuclear fragmentation in \noun{QGSJET}  is strongly correlated 
with the interaction treatment; it is assumed that the excitation of the
nuclear spectator part is proportional to the number of inelastically 
``wounded'' nucleons. In case of comparatively low excitations, corresponding
to peripheral interactions, the  de-excitation proceeds via an evaporation of
 a certain number of nucleons or alpha-particles. Highly excited nuclei
undergo  multi-fragmentation, which is described as a percolation process:
final fragments are formed by the clusters of nearby nucleons \cite{kal93}.
The scheme provides a reasonably good description of  the mass yield dependence for all
fragment masses, including intermediate ones, which is typically not the case if 
 evaporation mechanism only is included (e.g., in \cite{eng92}).
 Importantly, the relevant model parameters have been fixed
with the data on nuclear fragmentation at energies above 2 AGeV, where the
measured fragment yields are already stable and do not have any significant
energy dependence. Correspondingly, the treatment can be safely extrapolated
to cosmic ray energies.

\section{DISCUSSION OF MODEL RESULTS} \label{results}
\subsection{Cross sections and inelasticity}
The longitudinal air shower development depends on a number of characteristics
of hadron-air interactions. While the largest effect comes from the inelastic
proton-air (nucleus-air) cross section, which defines the position of the first
interaction of the primary particle, pion-air and kaon-air cross sections and
the inelasticities of hadronic interactions influence the shape of the 
characteristic EAS profile. In particular, cross sections of diffraction
dissociation processes play here a special role. The diffraction of the 
target nucleus results in a negligible energy loss of the initial particle,
 secondary hadrons being produced in the target fragmentation region,
and is essentially equivalent to a  reduction of the inelastic cross section.
The projectile hadron diffraction, being predominantly a coherent process,
the target nucleus remaining  intact with a high probability \cite{kal93},
results in a  small energy loss
of the leading particle, thus sizably influencing the overall inelasticity.
It is noteworthy, that all the above-mentioned quantities mainly depend 
on the model description of hadronic collisions in the
 peripheral and ``transition'' regions. Indeed, at small impact parameters, 
hadrons already look ``black'', see Fig.~\ref{fig:hadron-profile}~(d), with 
the interaction profile $\sigma ^{\rm inel}_{h-{\rm air}}(s,b)$ 
(the probability of inelastic interaction at some  impact parameter $b$
for a given energy $s$) being very close to unity,
 which still remains so when non-linear corrections to parton dynamics,
e.g., parton saturation effects, are taken into account.
This region does not contribute substantially to diffraction cross sections,
as any rapidity gap, produced in some elementary   production process,
is  covered by secondary particles created in other production
sub-processes. Finally, the inelasticity can only weakly depend
on the central region treatment. Central collisions  involve a large
number of  production processes and lead in all cases 
to large energy losses of leading particles.
 Contrary to that, modifications of the  treatment of the
peripheral and the ``transition'' regions make a sizable effect on the
predicted  $\sigma ^{\rm inel}_{h-{\rm air}}(s,b)$ behavior, thus influencing 
the obtained values of $\sigma ^{\rm inel}_{h-{\rm air}}$ and 
$\sigma ^{\rm diffr}_{h-{\rm air}}$, and change the average number and 
fluctuations of the number of production sub-processes, which leads 
to large variations of the  inelasticity.

As already mentioned, the   generalization of the formalism, discussed in
Sections  \ref{pomform}--\ref{nonlinear}, for hadron-nucleus and
 nucleus-nucleus interactions proceeds in a parameter-free way, 
which applies also to  cross section calculations. For the latter, 
one obtains  an excellent agreement with experimental data,
 some values given in Table~\ref{sig-exp}.%
\begin{table}[t]
\begin{center}
\small
\caption{Inelastic and non-diffractive hadron-carbon cross sections (in mb)
at 200 GeV lab. energy.}
\label{sig-exp}
\renewcommand{\tabcolsep}{.25pc} % enlarge column spacing
\renewcommand{\arraystretch}{1.1} % enlarge line spacing
\begin{tabular}{lccc}
\hline
  & $\sigma ^{\rm inel}_{n-^{12}\!C}$   & $\sigma ^{\rm ND}_{p-^{12}\!C}$  &
   $\sigma ^{\rm ND}_{\pi^+-^{12}\!C}$    
  \\ \hline
      \noun{QGSJET-II} & 238  & 224 & 167  \\
 Experiment  & $237\pm 2$ \cite{rob79} &  $225\pm 7$ \cite{car79} &
  $171\pm 5$ \cite{car79}  \\
\hline
\end{tabular}
\end{center}
\vspace{-.8cm}
\end{table}
The  predictions of the \noun{SIBYLL 2.1} model \cite{fle94,eng99} 
for $\sigma^{\rm inel}_{h-{\rm air}}$ are close to the ones of \noun{QGSJET-II}
 in the collider  range but rise faster with energy. 
In principle, this may come from the fact
that non-linear effects are introduced in SIBYLL for
semihard processes only, being neglected for the ``soft'' component.
As the latter plays the crucial role in peripheral collisions, which largely
 define the cross section behavior, one may expect that \noun{SIBYLL}
 overestimates $\sigma ^{\rm inel}_{h-{\rm air}}$ in the very
high energy range. Nevertheless,
model predictions depend also on the assumptions concerning hadron form factors
(parton distributions in the transverse plane) and on the calibration
to proton-proton data, as discussed in \cite{eng03}. Present uncertainties
concerning the energy dependence of  $\sigma ^{\rm inel}_{h-{\rm air}}$
will be drastically reduced after the measurements of 
 $\sigma ^{\rm tot}_{pp}$  at the LHC.

Model results for the inelasticity $K^{\rm inel}_{h-{\rm air}}$ 
of hadron-air interactions are only
 constrained by fixed target accelerator measurements. Practically  in all 
models the inelasticity rises with energy, due to the increase of the number
of elementary production processes. However, the spread in the predicted
 $K^{\rm inel}_{h-{\rm air}}$ is rather large, around 20\% in the high
energy range. Additional constraints may come from
 the measurements of leading neutron spectra
in proton-proton interactions by the LHCf experiment \cite{kas07}.

\subsection{Muon component of air showers}
An important topic are model predictions for the CR muon component. 
Here one has to distinguish between the results for inclusive muon spectra 
and for  the muon content of air showers of a given primary energy.
 The former are dominated by single interactions of primary protons of 
 energies in average only an order
  of magnitude higher than the ones of the measured muons. 
Due to the  steepness of the primary CR spectrum,   the corresponding results
are very sensitive to the shape of the forward pion and kaon spectra 
in proton-air collisions. The characteristic quantities are
 the so-called $Z$-factors \cite{gaisser}, some values given in
 Table \ref{zfact}%
\begin{table}[t]
\begin{center}\small
\caption{Pion and kaon $Z$-factors as predicted by the   \noun{QGSJET-II} and \noun{SIBYLL 2.1} models.}
\label{zfact}
\renewcommand{\tabcolsep}{.5pc} % enlarge column spacing
\begin{tabular}{lcccc}
\hline
 $E_{0}$, TeV &  model & $Z_{\pi^{+}}$  & $Z_{\pi^{-}}$  
& $Z_{K^{+}}$  \\ \hline
0.1   &   \noun{QGSJET-II} & 0.043  & 0.035 & 0.0036 \\
 &   \noun{SIBYLL 2.1}  & 0.036  & 0.026  & 0.0134 \\ \hline
10    &  \noun{QGSJET-II}  & 0.033  & 0.028  & 0.0034 \\
 &   \noun{SIBYLL 2.1}   & 0.037  & 0.029  & 0.0097  \\
\hline
\end{tabular}\end{center}
\vspace{-.8cm}
\end{table}
for    \noun{QGSJET-II} and \noun{SIBYLL 2.1}. Clearly,
 the two models predict  different energy dependences
for $Z_{\pi^{\pm}}$: one observes a rather precise Feynman scaling
in   \noun{SIBYLL}, which is supported by inclusive muon flux 
measurements \cite{ung05},
 and a noticeable scaling violation in 
 \noun{QGSJET-II}. Another difference is the higher values of $Z_{K^{+}}$
in case of \noun{SIBYLL}, which is due to the harder kaon spectra in the model.

The EAS muon content is formed
 during a multi-step hadronic cascade process and mainly depends on the
 total  multiplicity of hadron-air collisions,
the shape of the forward pion spectra being of secondary importance.
Model calibration is mainly performed at fixed target energies; in particular,
both \noun{QGSJET-II} and \noun{SIBYLL 2.1} appear to be consistent
 with  recent  data of the NA49 Collaboration \cite{na49}.
Concerning the predicted multiplicity,
 models agree with each other in the collider  range;
  at the highest CR energies the characteristic differences reach a factor of 
three. Most likely, the situation will not improve significantly
with the start of the  LHC, as the spread in model results
 for the multiplicity of proton-proton interactions
is yet at 10\% level in the corresponding energy range.

However, one should not expect large differences between model predictions
for the EAS muon number $N_{\mu}$. In particular, 
 \noun{QGSJET-II} and  \noun{SIBYLL 2.1} results for $N_{\mu}$ differ
 by only $10\div15$\% in the high energy range. 
As shown in \cite{ost06d}, changing the
multiplicity of proton-air collisions by a factor of two, one obtains less than
10\% modification of the predicted $N_{\mu}$. Of course,  pion-air
 multiplicity is of greater importance here. 
However, in \noun{QGSJET-II} one has
little freedom to modify the latter. The only new parameters, defining the
transition from proton-proton to pion-proton case, are    the residue $\gamma _{\pi}$ and the slope $R^2_{\pi}$ of the  Pomeron-pion coupling, which are
reliably fixed by data on pion-proton total cross section and elastic scattering
slope. The low $x$ behavior of the gluon and sea quark momentum distributions
in pion are described by the same soft Pomeron asymptotics as in the proton
case and the corresponding normalization is defined by the momentum
sum rule, the valence quark distributions being fixed by measurements.
It is therefore surprising that the new \noun{EPOS} model \cite{wer06}
predicts significantly higher $N_{\mu}$, which is almost a factor of two 
in access of the \noun{QGSJET-II} results at the highest CR energies
 \cite{wer06a}. Most likely, this is connected to the
  treatment of pion-air interactions in \noun{EPOS}. The picture
 can be validated by the KASCADE experiment,
 where the muon component is measured directly,  the 
corresponding analysis being sensitive to 10\% variations in the predicted
muon number \cite{ant05}.

\vspace{1mm}
{\small The author acknowledges fruitful discussions with H.J.~Drescher,
R.~Engel, and M.~Strikman.}

\end{document}